\documentclass[prb,twocolumn,showpacs,preprintnumbers,amsmath,amssymb]{revtex4}

\usepackage{graphicx}
\usepackage{dcolumn}
\usepackage{bm}

\begin{document}


\title{Two-dimensional incommensurate magnetic fluctuations in Sr$_2$(Ru$_{0.99}$Ti$_{0.01}$)O$_4$}

\author{K.~Iida$^1$}
\author{J.~Lee$^{1,2}$}
\author{M.~B.~Stone$^2$}
\author{M.~Kofu$^1$}\altaffiliation{Present address: Neutron Science Laboratory, Institute for Solid State Physics, University of Tokyo, Kashiwa, Chiba 277-8581, Japan.}
\author{Y.~Yoshida$^3$}
\author{S.-H.~Lee$^1$}\email{shlee@virginia.edu}

\affiliation{$^1$Department of Physics, University of Virginia, Charlottesville, Virginia 22904, USA}
\affiliation{$^2$Quantum Condensed Matter Division, Oak Ridge National Laboratory, Oak Ridge, Tennessee 37831, USA}
\affiliation{$^3$National Institute of Advanced Industrial Science and Technology, Tsukuba, Ibaraki 305-8565, Japan}

\date{\today}

\begin{abstract}
We investigate the imaginary part of the wave vector dependent dynamic spin susceptibility in Sr$_2$(Ru$_{0.99}$Ti$_{0.01}$)O$_4$ as a function of temperature using neutron scattering.
At $T=5$~K, two-dimensional incommensurate (IC) magnetic fluctuations are clearly observed around $\mathbf{Q}_\text{c}=(0.3,0.3,L)$ up to approximately 60~meV energy transfer.
We find that the IC excitations disperse to ridges around the $(\pi,\pi)$ point.
Below 50~K, the energy and temperature dependent excitations are well described by the phenomenological response function for a Fermi liquid system with a characteristic energy of 4.0(1)~meV.
Although the wave vector dependence of the IC magnetic fluctuations in Sr$_2$(Ru$_{0.99}$Ti$_{0.01}$)O$_4$ is similar to that in the Fermi liquid state of the parent compound, Sr$_2$RuO$_4$, the magnetic fluctuations are clearly suppressed by the Ti-doping.
\end{abstract}

\maketitle

Superconductivity in the Sr$_2$RuO$_4$ ruthenate has been studied for well over a decade in spite of its low transition temperature of $T_\text{c}=1.5$~K.\cite{Discover,Review}
This is because Sr$_2$RuO$_4$ has unconventional properties: this system has the same crystal structure as La$_{2-x}$Ba$_x$CuO$_4$, novel $p$-wave spin-triplet superconductivity,\cite{pWave,pWave2,NMR,Triplet0} and spontaneous time-reversal symmetry breaking.\cite{muSR,Kerr}
The vector order parameter is proposed to be $\hat{p}_x+i\hat{p}_y$,\cite{NMR,Triplet} but the coupling mechanism is still under debate.

The normal state of the Sr$_2$RuO$_4$ ruthenate may shed light on understanding the nature of the superconducting phase.
Both the electronic structure and imaginary part of the generalized magnetic susceptibility ($\chi''$) of this system have been studied intensely.
The $t_{2g}$ electrons of the Ru$^{4+}$ ions form three bands near the Fermi surface.
The $d_{xz}$ and $d_{yz}$ orbitals form quasi-one-dimensional $\alpha$ and $\beta$ sheets, while $d_{xy}$ forms a two-dimensional $\gamma$ sheet.\cite{BandCal0,BandCal,ARPES}
This low-dimensional band structure is one explanation for two-dimensional Fermi liquid behavior in Sr$_2$RuO$_4$.\cite{2DFermi}
The two-dimensional Fermi liquid behavior also manifests itself in the observation of incommensurate (IC) magnetic fluctuations up to 80~meV at the wave vector $\mathbf{Q}_\text{c}=(0.3,0.3,L)$ in Sr$_2$RuO$_4$.\cite{Neutron0,Q2D,Neutron1,Neutron2,Neutron3,Neutron4,Iida}
The IC fluctuations are due to Fermi surface nesting of the $\alpha$ and $\beta$ bands.\cite{Theory1,vanHove,Theory3,Gamma,Theory2,Theory0}
In order to fully understand the role of the magnetic fluctuations in Sr$_2$RuO$_4$, it is important to examine the system without the complication of the superconducting phase.
We substitute a small percentage of nonmagnetic ions for the Ru ions to access this phase.

Substitution of nonmagnetic Ti$^{4+}$ ions for Ru$^{4+}$ in Sr$_2$RuO$_4$ has already been examined.
Resistivity,\cite{Doping1,Doping4} magnetization,\cite{Doping1,Doping4} optical,\cite{Doping4} heat capacity,\cite{Doping4,Doping2} Sr$^{87}$ NMR,\cite{Doping3} and neutron scattering~\cite{OrderedState} measurements all report a significant doping effect in Sr$_2$(Ru$_{1-x}$Ti$_x$)O$_4$.
Sr$_2$(Ru$_{1-x}$Ti$_x$)O$_4$ shows superconducting behavior for $x\le0.0015$, an IC magnetic ordering or spin density wave state for $x\ge0.025$, and paramagnetic behavior for the intermediate substitution range $0.0015<x<0.025$.
In the small $x$ region (at least up to $x=0.19$), the space group is the same as the parent compound $I4/mmm$.\cite{Doping6}
Sr$_2$(Ru$_{0.91}$Ti$_{0.09}$)O$_4$ shows magnetic ordering at $T_\text{N}=25$~K at the IC wave vector $\mathbf{Q}_\text{IC}=(0.307,0.307,1)$.
This IC ordering is also due to Fermi surface nesting, and NMR measurements suggest that the Ti-doping enhances the IC fluctuation.\cite{Doping3}
In the intermediate region, Sr$_2$(Ru$_{1-x}$Ti$_x$)O$_4$ shows paramagnetic and two-dimensional Fermi liquid behavior.\cite{Doping1,Doping2}
However, to the best of our knowledge, there are no measurements or theoretical description of the extent of the magnetic fluctuations in the intermediate substitution region.

\begin{figure}[t]
\includegraphics[width=8.4cm,height=9.8cm]{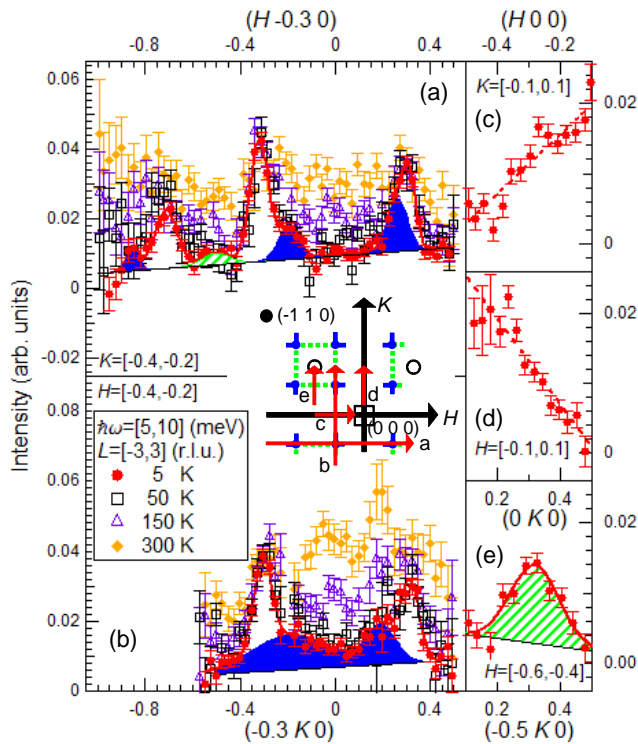}
\centering
\caption{(Color online).
Several constant-$\hbar\omega$ cuts of the neutron scattering intensities at $T=5$, 50, 150, and 300~K along five different directions in reciprocal space as described by arrows in the inset: along (a) $(H\ -0.3\ 0)$, (b) $(-0.3\ K\ 0)$, (c) $(H00)$, (d) $(0K0)$, and (e) $(-0.5\ K\ 0)$.
The energy window was from $\hbar\omega=4$ to 12~meV, and the incident neutron energy was $E_\text{i}=15$~meV.
Solid areas (blue) represent the shoulder peaks, and striped areas (green) the ridge intensities.
Solid lines (red) are fitting results at 5~K using a combination of Gaussians and a linear background.
Dashed lines (red) are guide to the eye.
The inset describes the schematic view of the magnetic fluctuation in the ($HK0$) plane.
Big solid circles (black) represent the nuclear Bragg reflections, while open circles and square are the $(\pi,\pi)$ and $\Gamma$ point, respectively.
Small solid circles (blue) describe the IC magnetic peaks, thin lines (blue) the positions of the shoulder peaks, and dashed lines (green) the ridge intensities between IC peaks.
Five thin arrows (red) show the directions in the reciprocal plane for each constant-$\hbar\omega$ cut shown in the main panels.
}\label{Fig:HKdependence}
\end{figure}

In this paper, we report measurements of the magnetic fluctuations in the intermediate substitution region of Sr$_2$(Ru$_{1-x}$Ti$_x$)O$_4$ ($x=0.01$) using time-of-flight inelastic neutron scattering.
Sr$_2$(Ru$_{0.99}$Ti$_{0.01}$)O$_4$ exhibits two-dimensional IC magnetic fluctuations around $\mathbf{Q}_\text{c}=(0.3,0.3,L)$, and a ridge scattering was observed around the $(\pi,\pi)$ position rather than the $\Gamma$ point at 5~K; the positions of IC magnetic fluctuations are almost the same as in Sr$_2$RuO$_4$.\cite{Iida}
The IC peak, shoulder peak, and ridge scattering positions in the reciprocal lattice are summarized in the inset of Fig.~\ref{Fig:HKdependence}.
IC magnetic fluctuations were observed up to 60~meV, and the characteristic energy of the spin fluctuation, $\hbar\omega_\text{SF}$, are estimated as $\hbar\omega_\text{SF}=4.0(1)$~meV.
The value of $\hbar\omega_\text{SF}$ is slightly smaller than that in Sr$_2$RuO$_4$,\cite{Iida} indicating that the Ti-doping suppresses the IC magnetic fluctuations.
The temperature dependence of the magnetic fluctuations indicates that Fermi liquid behavior in Sr$_2$(Ru$_{0.99}$Ti$_{0.01}$)O$_4$ is only present below $T=50$~K which is close to $\hbar\omega_\text{SF}/k_\text{B}\sim46(2)$~K.

Three single crystals of Sr$_2$(Ru$_{0.99}$Ti$_{0.01}$)O$_4$ with a total mass of $\sim11$~g were prepared by the floating-zone method.\cite{Sample1,Sample2}
SQUID magnetometer magnetization measurements down to 0.3~K observed no superconductivity.
Neutron scattering measurements were done at the wide angular-range chopper spectrometer, ARCS at SNS.\cite{ARCS}
Neutrons with incident energies of $E_\text{i}=15$, 80, and 150~meV were scattered from the sample which was aligned in the $(HK0)$ scattering plane.
The sample was oriented with the $(00L)$ vector parallel to the incident neutrons.
The sample was contained within an aluminum can with a He exchange gas.
The sample can was attached to a closed-cycle displex refrigerator.
Measurements with $E_\text{i}=15$~meV were performed at 5, 50, 150, and 300~K, and measurements with $E_\text{i}=80$ and 150~meV were performed at 5~K.
Background measurements for $E_\text{i}=15$~meV were performed by measuring an empty sample can at each temperature.
This empty can background was subtracted from all $E_\text{i}=15$~meV measurements.

\begin{figure}[t]
\includegraphics[width=8.54cm,height=9.8cm]{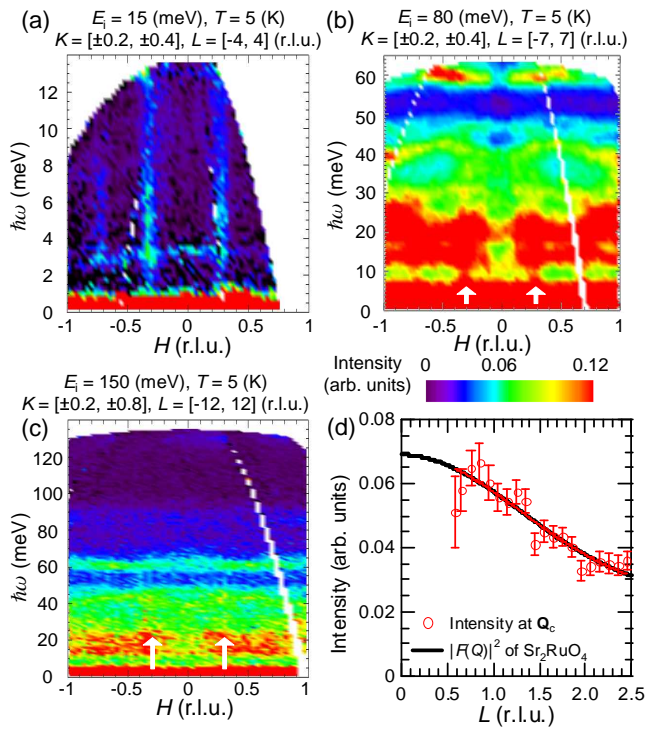}
\centering
\caption{(Color online).
(a)--(c) Contour maps of the neutron scattering intensity as a function of $H$ and $\hbar\omega$ at $|K|=0.3$ [$|K|=0.5$ for (c)], $L=0$, and $T=5$~K; the incident energies of neutrons were $E_\text{i}=15$, 100, and 150~meV, respectively.
Data have been integrated over $K$ and $L$ as noted at the top of each panel.
Arrows (white) in panels (b) and (c) represent the IC peak positions for the high energy regions, and (b) is shown at 1.25 times the intensity scale of the color bar.
(d) $L$-dependence of the neutron scattering intensity at $H=K=[\pm0.25,\pm0.35]$, $\hbar\omega=[2.5,10]$~meV, and $T=5$~K with $E_\text{i}=15$~meV (circles) and the squared magnetic form factor of Sr$_2$RuO$_4$~\cite{Neutron4} (line).
}\label{Fig:Hwdependence}
\end{figure}

Figures~\ref{Fig:Hwdependence}(a)--\ref{Fig:Hwdependence}(c) show neutron scattering intensities along the $H$-direction for energy transfers up to $\hbar\omega=130$~meV.
Strong scattering intensities from phonons are observed around $\hbar\omega=40$, 60, and 85~meV.
These are at very similar values as observed in Sr$_2$RuO$_4$.\cite{Phonon,Iida}
The magnetic fluctuations can be seen at low energies where phonon scattering is relatively week for small values of $|\mathbf{Q}|$.
Figure~\ref{Fig:Hwdependence}(a) clearly shows strong spin fluctuations at the IC positions centered at $\mathbf{Q}_\text{c}=(0.3,0.3)$.
As shown in Figs.~\ref{Fig:Hwdependence}(b) and \ref{Fig:Hwdependence}(c), we observe the IC magnetic fluctuations to extend up to $\hbar\omega\approx60$~meV, with no clear intensity beyond that point.
We also find that the $L$-dependence of the IC peak shows agreement with the squared magnetic form factor of Sr$_2$RuO$_4$~\cite{Neutron4} as shown in Fig.~\ref{Fig:Hwdependence}(d).
This suggests that the IC magnetic fluctuation in Sr$_2$(Ru$_{0.99}$Ti$_{0.01}$)O$_4$ is two-dimensional, just as found in Sr$_2$RuO$_4$.\cite{Q2D,Neutron1,Neutron4}

\begin{figure}[t]
\includegraphics[width=7.56cm,height=4.2cm]{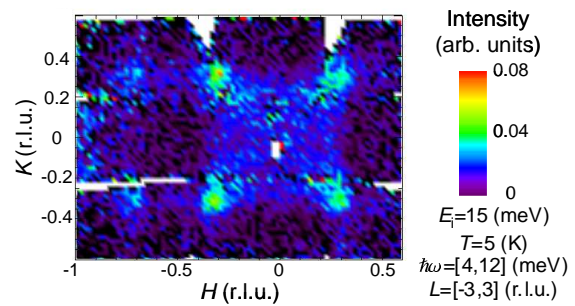}
\centering
\caption{(Color online).
Contour map of a constant-$\hbar\omega$ slice of the neutron scattering intensity at $T=5$~K with an incident neutron energy of $E_\text{i}=15$~meV.
Energy transfer was integrated from $\hbar\omega=4$ to 12~meV, and $L$ was integrated from $-3$ to $3$.
}\label{Fig:Mapping}
\end{figure}

A contour map of the neutron scattering intensity as a function of $\mathbf{Q}=(H,K)$ is shown in Fig.~\ref{Fig:Mapping}.
$S(\mathbf{Q},\hbar\omega)$ at $T=5$~K was integrated over $\hbar\omega$ from 3 to 10~meV and $L$ from $-2.6$ to 2.6.
Several IC peaks with the characteristic wave vector of $\mathbf{Q}_\text{c}=(0.3,0.3)$ are observed.
This is very similar to what is observed in Sr$_2$RuO$_4$.\cite{Iida}
Alternatively, no sizable peak was observed in the contour map in the $(HK0)$ plane within the elastic channel (not shown).

In order to further characterize the magnetic excitations, we integrate the neutron scattering intensity in the low energy region along several different $\mathbf{Q}$-directions as shown in the inset of Fig.~\ref{Fig:HKdependence}.
The paths labeled in this inset correspond to the different panels of Fig.~\ref{Fig:HKdependence}.
Figures~\ref{Fig:HKdependence}(a) and \ref{Fig:HKdependence}(b) show the IC peaks at several temperatures.
Using a combination of Gaussians and a linear background, the IC peak positions at $T=5$~K are $H=-0.711(6)$, $-0.319(3)$, and $0.31(1)$ with $K\approx-0.3$ and $K=-0.304(3)$ and $0.316(8)$ with $H\approx-0.3$.
In addition, there are some shoulder peaks shown by solid areas (blue) of the fitted curve at $H=-0.85(1)$, $-0.20(1)$, and $0.26(7)$ with $K\approx-0.3$ and $K=-0.20(4)$ and $0.19(3)$ with $H\approx-0.3$.
For the possible ridge scattering, we plot the data around the $\Gamma$ point [see the inset of Fig.~\ref{Fig:HKdependence} for the directions and Figs.~\ref{Fig:HKdependence}(c) and \ref{Fig:HKdependence}(d) for data] and across the ridges around $(\pi,\pi)$ [Fig.~\ref{Fig:HKdependence}(e)].
The data show significant ridge scattering around the $(\pi,\pi)$ point [peak at $K=0.32(1)$ with $H\approx-0.5$ in Fig.~\ref{Fig:HKdependence}(e)] rather than around the $\Gamma$ point [no clear peak in Figs.~\ref{Fig:HKdependence}(c) and \ref{Fig:HKdependence}(d)].\cite{Theory1,Theory0,Neutron1,Iida}
This can also be seen in the data of Fig.~\ref{Fig:HKdependence}(a).
There is extra scattering intensity between the peaks at $H=-0.7$ and $-0.3$ represented by dashed areas (green) in Fig.~\ref{Fig:HKdependence}(a), but there is no additional peak in the scattering intensity between $H=-0.3$ and $0.3$ except for the shoulder intensities (blue shaded region).

As shown in Figs.~\ref{Fig:HKdependence}(a) and \ref{Fig:HKdependence}(b), the IC fluctuations at 50~K are at the same location as at 5~K.
At 150~K, the IC fluctuation peaks becomes broader, but still clearly observed.
Unlike Sr$_2$RuO$_4$,\cite{Iida} however, the IC magnetic fluctuation becomes almost flat at 300~K.

\begin{figure}[t]
\includegraphics[width=8.4cm,height=11.2cm]{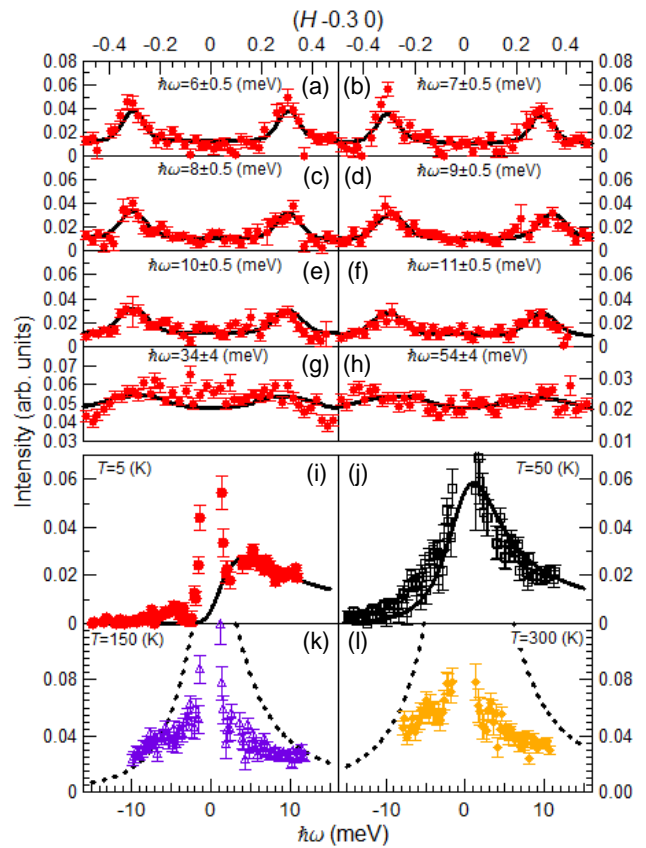}
\centering
\caption{(Color online).
$H$-dependences of constant-$\hbar\omega$ cuts at $T=5$~K with the energy windows of (a) $\hbar\omega=6\pm0.5$, (b) $7\pm0.5$, (c) $8\pm0.5$, (d) $9\pm0.5$, (e) $10\pm0.5$, (f) $11\pm0.5$, (g) $34\pm4$, and (h) $54\pm4$~meV.
$K$ and $L$ were integrated from $-0.4$ to $-0.2$ and from $-9$ to $9$, respectively.
This cut is path a as shown in the inset of Fig.~\ref{Fig:HKdependence}.
The incident neutron energies were (a)--(f) $E_\text{i}=15$, and (g) and (h) 80~meV.
Fitting results using Eq.~(\ref{Eq:SatoMaki}) with a linear background are shown as solid lines in each panel.
$\hbar\omega$-dependences of the IC fluctuation at (i) $T=5$, (j) 50, (k) 150, and (l) 300~K with $E_\text{i}=15$~meV.
$H$, $K$, and $L$ were integrated from $|\pm0.25|$ to $|\pm0.4|$, from $|\pm0.25|$ to $|\pm0.4|$, and from $-9$ to $9$, respectively.
Calculated results using Eq.~(\ref{Eq:SatoMaki}) ignoring the background at the elastic positions are described by solid lines in panels (i) and (j) [dashed lines in panels (k) and (l)].
Solid lines are a simultaneous fit to the $T\le50$~K data as described in the text.
}\label{Fig:SatoMaki}
\end{figure}

We also examine the $H$-dependences of the IC magnetic fluctuations at 5~K in several different $\hbar\omega$ windows as shown in Figs.~\ref{Fig:SatoMaki}(a)--\ref{Fig:SatoMaki}(h).
Our data clearly shows that IC magnetic fluctuations in Sr$_2$(Ru$_{0.99}$Ti$_{0.01}$)O$_4$ survive all the way up to $\sim60$~meV, and the IC peak position does not change even in the high $\hbar\omega$ region.
Figures~\ref{Fig:SatoMaki}(i)--\ref{Fig:SatoMaki}(l) show $\hbar\omega$-dependences at $\mathbf{Q}_\text{c}$ for several temperatures.
The maximum intensity of the IC magnetic fluctuation is centered around $\hbar\omega=5$~meV at 5~K.
At 50 and 150~K, magnetic fluctuations on the neutron energy loss side are merged into the elastic positions, and the scattering on the neutron energy gain side begins to increase due to detailed balance.
At 300~K, the magnetic fluctuation becomes symmetric centered at the elastic position.
We explain the $\hbar\omega$ and $T$ dependent scattering of the IC magnetic fluctuations using the response function of a Fermi liquid.

To analyze the energy dependent IC magnetic excitations, we compare the data to the general form of the phenomenological response function used to describe a Fermi liquid system~\cite{Hayden} with the fluctuation dissipation theorem:
\begin{eqnarray}\label{Eq:SatoMaki}
S(\mathbf{Q},\hbar\omega)&=&\frac{\left|F(Q)\right|^2}{1-\text{exp}(-\hbar\omega/k_\text{B}T)}\\
&\times&\sum\limits_{\mathbf{Q}_\text{c}}\frac{\chi_\delta\kappa_0^4\left(\hbar\omega/\hbar\omega_\text{SF}\right)}{\left[\kappa_0^2+\left(\mathbf{Q}-\mathbf{Q}_\text{c}\right)^2\right]^2+\left(\hbar\omega/\hbar\omega_\text{SF}\right)^2\kappa_0^4}\nonumber
\end{eqnarray}
where $\chi_\delta$, $\kappa_0$, $\hbar\omega_\text{SF}$, and $\mathbf{Q}_\text{c}$ are parameters for the peak intensity, the sharpness of the peak, the characteristic energy of the spin fluctuations, and the IC peak position.\cite{SatoMaki1,SatoMaki2}
These variables are all independent of $\mathbf{Q}$ and $\hbar\omega$.
The magnetic form factor of Sr$_2$(Ru$_{0.99}$Ti$_{0.01}$)O$_4$, $F(Q)$, was obtained from that of Sr$_2$RuO$_4$.\cite{Neutron4}
As shown in Figs.~\ref{Fig:SatoMaki}(a)--\ref{Fig:SatoMaki}(j), we fit the $H$- and $\hbar\omega$-dependences at each temperatures (5 and 50~K) simultaneously to Eq.~(\ref{Eq:SatoMaki}) with a simple linear background.
The elastic positions in Figs.~\ref{Fig:SatoMaki}(i) and \ref{Fig:SatoMaki}(j) are ignored in the fitting process.
From fitting only the $\hbar\omega=6$~meV data [Fig.\ref{Fig:SatoMaki}(a)], $\mathbf{Q}_\text{c}$ was obtained to be $\mathbf{Q}_\text{c}=(0.308(4),0.310(4))$, and fixed for the global fit.
As shown by the solid lines, Eq.~(\ref{Eq:SatoMaki}) describes our data well over the entire energy range below 50~K.
The optimum parameters obtained from the best fit are $\chi_\delta=0.100(2)$, $\kappa_0=0.048(2)$~r.l.u. [$=0.078(4)$~$\text{\AA}$$^{-1}$] and $\hbar\omega_\text{SF}=4.0(1)$~meV.
In addition, the calculated intensities using Eq.~(\ref{Eq:SatoMaki}) with obtained parameters at 150 and 300~K are also shown in Figs.~\ref{Fig:SatoMaki}(k) and \ref{Fig:SatoMaki}(l) as dashed lines.
There are apparent discrepancies, indicating that Eq.~(\ref{Eq:SatoMaki}) can only describe this system below 50~K which is close to the characteristic energy of $\hbar\omega_\text{SF}/k_\text{B}=46(2)$~K.
Above this temperature thermal population of states and additional phonon excitations makes it difficult to extract only the magnetic excitations from the measurement.

By comparing the observed IC magnetic fluctuations in Sr$_2$(Ru$_{0.99}$Ti$_{0.01}$)O$_4$ to those in the parent compound ruthenate, the IC peak positions are almost identical, but the value of $\hbar\omega_\text{SF}$ is smaller [$\hbar\omega_\text{SF}=5.0(3)$~meV in Sr$_2$RuO$_4$~\cite{Iida}].
This indicates that Ti-doping suppresses the IC magnetic fluctuations, although an NMR study reported that the Ti-doping enhances the IC magnetic fluctuation in Sr$_2$(Ru$_{1-x}$Ti$_{x}$)O$_4$.\cite{Doping3}

In summary, our inelastic neutron scattering measurements of Sr$_2$(Ru$_{0.99}$Ti$_{0.01}$)O$_4$ over a very wide range of wave-vector transfer, energy transfer and temperature reveal the three components of the two-dimensional magnetic fluctuations at 5~K: strong IC spin fluctuations centered at $\mathbf{Q}_\text{c}=(0.3,0.3,L)$ that extend up to approximately 60~meV with a characteristic energy of $\hbar\omega_\text{SF}=4.0(1)$~meV, the weaker so-called shoulder scattering, and the ridge scattering in the vicinity of $\mathbf{Q}_\text{c}$ at low energy transfer.
Our data clearly show that the ridge scattering is more significant near the $(\pi,\pi)$ wave vector rather than around the $\Gamma$ points.
Below 50~K, energy and temperature dependences are well described by the phenomenological response function for the Fermi liquid system.
Although the IC peak positions in Sr$_2$(Ru$_{0.99}$Ti$_{0.01}$)O$_4$ are the same as those in the parent compound ruthenate, the magnetic fluctuation spectrum is clearly suppressed.

We would like to thank M. Matsuda for his help in our neutron scattering experiments.
This work at the University of Virginia was supported by the US NSF under Agreement No.~DMR-0903977.
The Research at Oak Ridge National Laboratory's Spallation Neutron Source was sponsored by the Scientific User Facilities Division, Office of Basic Energy Sciences, U. S. Department of Energy.

\end{document}